# Coherent Phonon Control of Ultrafast Magnetization Dynamics in $Fe_3GeTe_2$ from Time-Dependent *Ab Initio* Theory

Zhaobo Zhou[1], Min Li[1], Thomas Frauenheim[2,3], Junjie He[1,*]

[1]Department of Physical and Macromolecular Chemistry, Faculty of Science, Charles University in Prague, Prague 12843, Czech Republic
[2]School of Science, Constructor University, Bremen 28759, Germany
[3]Institute for Advanced Study, Chengdu University, Chengdu 610106, China
*E-mail: junjie.he@natur.cuni.cz

**Abstract**: Exploring ultrafast magnetization control in two-dimensional (2D) magnets through optically driven coherent phonons has been well-established. Yet, the microscopic interplay between spin dynamics and lattice degrees of freedom remains less explored. Employing real-time time-dependent density functional theory (rt-TDDFT) coupled with Ehrenfest dynamics, we systematically investigate laser-induced spin-nuclei dynamics with coherent phonon excitation in the 2D ferromagnet $Fe_3GeTe_2$. We found that selectively pre-exciting three typical coherent phonon modes results in up to a 53% additional spin moment loss in an out-of-plane $A_{1g}^2$ mode within ~50 fs. Coherent phonon control of spin dynamics is closely linked to laser pulse parameters. The underlying microscopic mechanism of this phenomenon is primarily governed by coherent phonon-induced asymmetric spin-resolved charge transfer following the disappearance of the laser pulse, thereby enabling effective control of the spin moment loss. Our findings offer a novel insight into the coupling of coherent phonons with spin systems in 2D limits on femtosecond timescales.

## 1. Introduction

The ultrafast manipulation of magnetization dynamics via laser pulses has emerged as a focal point within the realm of opto-spintronics in recent years, owing to its fast and low energy consumption.[1–5] The electromagnetic field inherent to laser pulses facilitates interactions with electronic degrees of freedom, eliciting charge excitations and indirectly coupling to spin,[6] thereby enabling regulation of magnetization dynamics. Nevertheless, the pivotal role played by laser-induced lattice vibrations in governing ultrafast magnetization dynamics often remains underappreciated. Fundamentally, the lattice acts as a reservoir for energy and momentum, into which the angular momentum lost during demagnetization is transferred.[7] This lattice-mediated mechanism can precipitate spin transfers between magnetic sublattices on an ultrafast time scale,[8,9] or potentiate exchange interactions over a longer time scale.[10] Thus, how the lattice degree of freedom comes into play holds paramount significance for the optical manipulation of magnetization dynamics.

Recently, the concept of phonomagnetism has garnered attention, wherein light is employed to manipulate the spin degree of freedom of magnetic ions by exciting coherent lattice vibrations, known as coherent phonons.[11–14] Diverging from traditional optomagnetism, phonomagnetism entails the transfer of angular momentum to magnetic ions[15–17] or the transient modulation of the lattice into a modified magnetic order.[18,19] This concept suggests a potential avenue for directly controlling magnetization dynamics via pre-excitation of coherent phonons preceding laser-induced spin dynamics. Various experimental methodologies, such as strain pulse,[20,21] impulsive stimulated Raman scatting,[22] and displacive excitation of coherent phonons,[23] have been developed to produce coherent phonons. The excitation of coherent phonons has also been instrumental in manipulating and probing optical-matter interactions, leading to the discovery of novel optical and physical phenomena.[21,24–28] However, theoretical investigations into utilizing coherent phonons for the ultrafast control of magnetization dynamics in magnetic materials, particularly in the context of two-dimensional (2D) magnets, remain scarce. Moreover, the underlying physical mechanisms governing coherent phonon-dependent magnetization dynamics processes remain ambiguous. Addressing this gap represents an urgent imperative, albeit it poses a significant challenge until two fundamental prerequisites are met. Firstly, identifying suitable materials exhibiting Raman-active coherent lattice vibrations that can be readily excited by laser pumps is crucial. Secondly, developing robust theoretical frameworks capable of effectively simulating the coupling between charge/spin dynamics and atomic nuclei motion is essential.

The treatment of the spin and charge dynamics for extended solid systems via real-time time-dependent density functional theory (rt-TDDFT) is now becoming a well-established theoretical framework. Notably, Dewhurst et al. utilized rt-TDDFT to introduce the optical intersite spin transfer (OISTR) effect, demonstrating that laser pulses can efficiently and coherently redistribute spins between distinct magnetic sublattices in magnetic materials.[8,29] Several experimental results have also confirmed the OISTR effect in various magnetic systems.[5,9,30] Employing rt-TDDFT, we also systematically reported the laser pulse-induced spin dynamics in 2D magnets and van der Waals heterostructures.[31–35] However, these works primarily focus on the pure electron dynamics triggered by laser pulses, neglecting the influence of nuclei. The coupling of the spin dynamics with nuclei poses a challenge in extended solids and remains a less-explored aspect of rt-TDDFT.[13]

In this work, using rt-TDDFT coupled with Ehrenfest dynamics, we systematically investigate laser-induced spin-nuclei dynamics with coherent phonon excitation in the 2D ferromagnet $Fe_3GeTe_2$ (FGT). Our simulations reveal that considering the motion of atomic nuclei leads to an additional loss of spin moment for Fe, Ge, and Te atoms. Especially for FGT with pre-excited out-of-plane $A_{1g}^2$ coherent phonons, the increment in spin moment loss of Fe atom reaches up to 53%. This phenomenon can be attributed to the nuclei motion of coherent phonon-induced asymmetric interatomic charge transfer subsequent to the cessation of the laser pulse. Moreover, when multiple coherent phonons are simultaneously pre-excited, the mode that induces a significant increase in spin moment loss predominates the magnetization dynamics. We also observe that the magnetization dynamics of FGT are susceptible to variations in laser pulse fluence, yet remain insensitive to changes in the amplitude of pre-excited coherent phonons.

## 2. Results and Discussion

To elucidate the impact of pre-excited coherent phonons on magnetization dynamics, we initially characterize the phonon spectra of the FGT monolayer, as depicted in **Figure 1**. Previous works reported that Raman active two $A_{1g}$ and one $E_{2g}$ coherent phonons at Γ point in FGT few-layer or flakes are observed at room temperature.[36–38] With a reduction in the number of layers, these Raman peaks typically undergo an upshift. Such observed phonon modes in previous experiments are also obtained (labeled as cyan stars) in our calculated phonon spectra (Figure 1a), which are well in agreement with that in previous work,[39] as shown in the inset panel in Figure 1a. Consequently, we consider three prominent phonon modes, namely $A_{1g}^1$, $A_{1g}^2$ and $E_{2g}$ modes, for pre-excitation in subsequent TDDFT calculations (Figures 1b-1d). Additionally, the $E_{2g}$ mode involves the in-plane vibrations of Fe, Ge and Te atoms. To further analyze which vibration component determines the magnetization dynamics of FGT with pre-excited $E_{2g}$ mode, we individually consider three in-plane modes: in-plane Fe, in-plane Ge, and in-plane Te modes (Figures 1e-1g). The double-pump method can be implemented experimentally, where these phonons are first pre-excited, followed by the application of an optical laser pulse to manipulate the spin dynamics while the phonon modes remain excited. Taking Fe atoms as an example, the atomic displacement of Fe1 and Fe2 atoms under the linearly polarized pulse with a photon energy of 1.55 eV, a 12.21 fs full width at half maximum (FWHM) and 11.20 mJ/cm$^2$ is shown in Figure S1 (Supporting Information). This pulse will be used in all calculations unless stated. It found that the Fe1 and Fe2 atoms exhibit a maximum out-of-plane phonon amplitude of 0.013 nm with a corresponding time period of 60 fs. Thus we set the pre-excited coherent phonons with an amplitude of 0.01 nm in the following calculation.

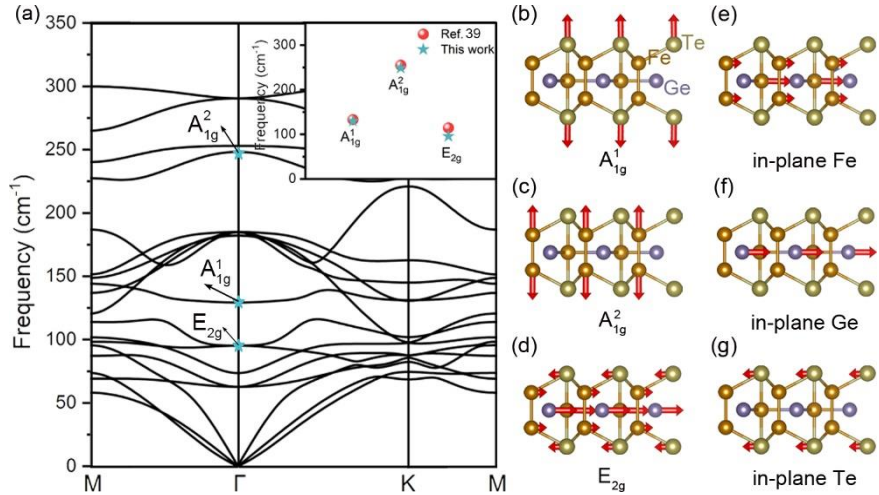

**Figure 1.** (a) Phonon spectra for the Fe$_3$GeTe$_2$ monolayer. The phonon frequency of two A$_{1g}$ modes and one E$_{2g}$

mode at Γ point are given in the inset panel, which is in agreement with the corresponding Raman shift in Ref. 39. (b-g) Two out-of-plane $A_{1g}$ modes ($A_{1g}^1$ and $A_{1g}^2$) and in-plane $E_{2g}$ mode, as well as the in-plane Fe, in-plane Ge and in-plane Te modes in $E_{2g}$ mode. The red arrows represent the amplitude vector.

After pre-exciting specific coherent phonons, we investigate the change in spin moment ($\Delta M(t)=M(t)-M(0)$) dynamics of Fe, Ge and Te atoms in FGT under the laser pulse excitation with full nuclear dynamics and in the absence of nuclear dynamics, as shown in **Figure 2** and Figure S2 (Supporting Information). Before 20 fs, the processes of moment loss for Fe, Ge, and Te atoms exhibit minimal discrepancy with and without nuclear dynamics. However, all the coherent phonons exhibit an additional spin moment loss with nuclear dynamics after 20 fs, revealing the significant effect of coherent phonons on manipulating magnetization dynamics in FGT. Particularly noteworthy is the pronounced effect observed for the out-of-plane $A_{1g}^2$ mode (Figure 2b), the spin moment loss increases by 1.2 Bohr magneton per Fe atom (53% of the ground-state moment). This augmentation can be attributed to nuclear dynamics encompassing forces arising from both the pre-excited phonon and momentum transfer from the excited electronic system. The latter contributes less than 5% of the total force and exerts a negligible influence on spin dynamics in the early stages.[13] Subsequently, as the pre-excited lattice returns to equilibrium under nuclear dynamics, the moving nuclei induce a robust back-reaction on the electronic system. These findings underscore the pivotal role of nuclear dynamics in decisively shaping ultrafast spin moment loss, thereby facilitating expedited spin manipulation. In addition, we observe a concurrent decrease in the spin moment of Fe atoms alongside a corresponding increase in the spin moment of Ge and Te atoms before 20 fs (Figure S2, Supporting Information), indicative of OISTR. However, the reversible impact of pre-excited coherent phonon-induced nuclei motion on the electronic system remains unclear. Consequently, we will delve deeper into the underlying physics of the coherent phonon-induced spin-resolved charge dynamics process in subsequent discussions.

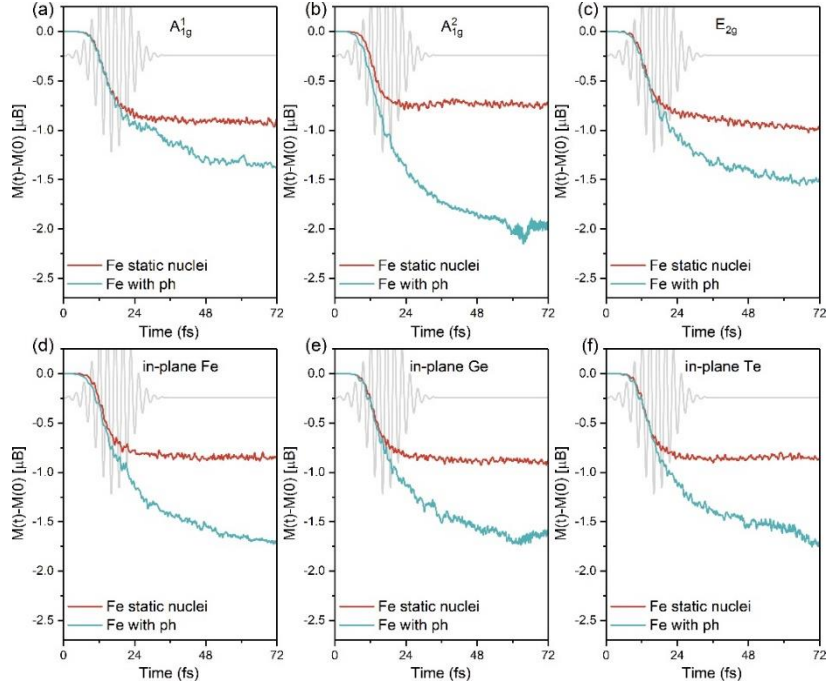

**Figure 2.** Change in the spin moment ($\Delta M(t)=M(t)-M(0)$) dynamics of Fe atom in FGT for full nuclear dynamics (Fe with ph) and in the absence of nuclear dynamics (Fe static nuclei). Results are shown for six pre-excited coherent

phonon modes: (a) $A_{1g}^1$, (b) $A_{1g}^2$, (c) $E_{2g}$, (d) in-plane Fe, (e) in-plane Ge and (f) in-plane Te.

To elucidate the underlying mechanism behind the additional spin moment loss induced by pre-excited coherent phonons, the dynamics of spin-resolved charge occupation of Fe atom in FGT upon pre-excited $A_{1g}^1$, $A_{1g}^2$ and $E_{2g}$ modes are examined and shown in **Figure 3**, which can be expressed as follows:

$$\Delta n_\uparrow = \frac{\Delta n(t) + \Delta M(t)}{2}, \Delta n_\downarrow = \frac{\Delta n(t) - \Delta M(t)}{2} \quad (1)$$

where $\Delta n(t) = n(t) - n(t=0)$ represents the change in local charge compared to the initial charge. $\Delta n_\uparrow$ and $\Delta n_\downarrow$ denote the time-dependent changes in spin-up and spin-down charges, respectively. Higher values indicate greater changes in charges. Figures 3b-3d illustrate that $\Delta n_\uparrow$ ($\Delta n_\downarrow$) of Fe decreases (increases) as a function of time, explaining the ultrafast demagnetization of Fe atoms in the absence of the nuclear dynamics of coherent phonons. However, the $\Delta n_\uparrow$ ($\Delta n_\downarrow$) of Fe atoms show more pronounced loss (gain) when nuclear dynamics are considered, indicating that the nuclear motion of coherent phonons amplifies the spin moment loss over a longer time scale. Nevertheless, the spin-resolved charge dynamics manifest greater complexity. Specifically, the $\Delta n_\uparrow$ and $\Delta n_\downarrow$ remain relatively stable for $A_{1g}^1$ and $E_{2g}$ modes, resulting in limited spin moment loss after ~22 fs (Figures 3b and 3d). Conversely, for the $A_{1g}^2$ mode, the $\Delta n_\uparrow$ of Fe atom continuously decreases while the $\Delta n_\downarrow$ increases before subsequently decreasing (Figure 3c). Such inequivalent alterations between $\Delta n_\uparrow$ and $\Delta n_\downarrow$ ultimately contribute to significant spin moment loss of Fe atoms upon $A_{1g}^2$ mode. To further underscore the influence of coherent phonons on magnetization dynamics, the difference between time-dependent charge with and without nuclear dynamics ($\Delta n_{ph-static}$) is depicted in Figures 3e-3g. As can be seen, it is evident that the spin-up $\Delta n_{ph-static}$ plays a predominant role in inducing substantial spin moment loss with $\Delta M_{A_{1g}^2} > \Delta M_{E_{2g}} > \Delta M_{A_{1g}^1}$.

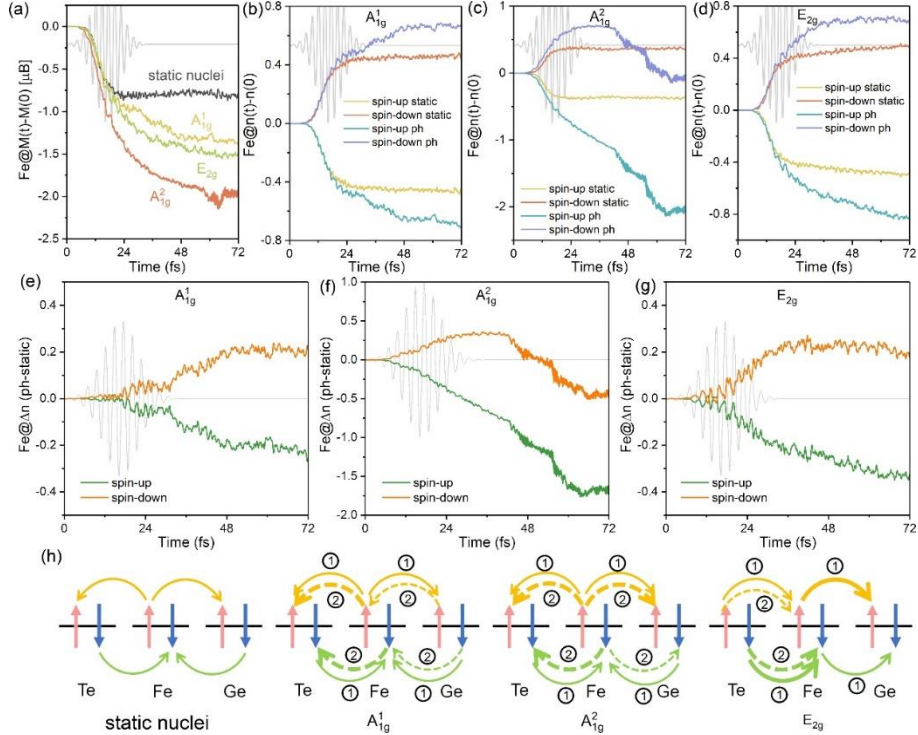

**Figure 3**. (a) Change in the spin moment ($\Delta M(t)=M(t)-M(0)$) dynamics of Fe atom in FGT for full nuclear dynamics

with pre-excited $A_{1g}^1$, $A_{1g}^2$ and $E_{2g}$ coherent phonons and in the absence of nuclear dynamics without pre-excited coherent phonons. (b-d) Change in the spin-resolved charge ($\Delta n(t)=n(t)-n(0)$) of Fe atom in FGT with (static) and without (ph) nuclear dynamics. Results are shown for pre-excited $A_{1g}^1$, $A_{1g}^2$ and $E_{2g}$ coherent phonons, respectively. The positive (negative) value means the increase (decrease) of electrons. (e-g) Spin-resolved charge difference of Fe atom between full nuclear dynamics and in the absence of nuclear dynamics with pre-excited $A_{1g}^1$, $A_{1g}^2$ and $E_{2g}$ coherent phonons, respectively. The positive (negative) value means the increase (decrease) of charge. (h) The schematic diagram of spin charge transfer dynamics for pre-excited $A_{1g}^1$, $A_{1g}^2$ and $E_{2g}$ coherent phonons with and without nuclear dynamics. The dash and dot arrows represent the charge transfer dynamics pathway dominated by laser and phonon, respectively. The thicker the arrow, the larger the amount of charge transfer.

As mentioned above, the presence of pre-excited coherent phonons with nuclear dynamics significantly influences the spin-resolved $\Delta n$ of Fe atoms, leading to additional spin moment loss. To give a clear physical description of the coherent phonon-induced spin-resolved charge dynamics process, we also examine the spin-resolved $\Delta n$ of Ge and Te atoms, as depicted in Figures S3-S5 (Supporting Information). Both Ge and Te atoms exhibit an increase in spin-up charge and a decrease in spin-down charge in the absence of nuclear dynamics. However, the spin-resolved $\Delta n$ displays asymmetry when nuclear dynamics are taken into account. For instance, the $\Delta n_\uparrow$ and $\Delta n_\downarrow$ of Te atoms undergo a sharp transition after ~48 fs, attributed to charge transfer from the adjacent Fe atoms. Based on these findings, the specific charge dynamics pathways in FGT for pre-excited $A_{1g}^1$, $A_{1g}^2$ and $E_{2g}$ coherent phonons with and without nuclear dynamics are illustrated in Figure 3h. The spin-up charge of Fe atoms will be transferred to that of Ge and Te atoms, while the spin-down charge of Fe will be gained from the Ge and Te atoms in FGT, in the absence of nuclear dynamics throughout the entire duration, a phenomenon attributed to the OISTR effect. However, the charge dynamics pathway can be divided into two parts when the nuclear dynamics is considered. First, the charge transfer between Fe and Ge/Te atoms occurs via OISTR (process ①), akin to scenarios without nuclear dynamics. After that, the nuclei motion of pre-excited $A_{1g}^1$, $A_{1g}^2$ and $E_{2g}$ coherent phonons governs the charge distribution between Fe and Ge/Te atoms, forming various coherent phonon-assisted charge transfer pathways over longer time scales (process ②). Note that the OISTR always exhibits competition with spin-flip mediated by spin-orbit coupling (SOC) in charge redistribution.[40] To estimate the SOC effect on the charge dynamics process, we calculate the spin-resolved charge difference of Fe atoms between full nuclear dynamics and in the absence of nuclear dynamics with and without SOC for pre-excited $A_{1g}^2$ coherent phonons, as illustrated in Figure S6 (Supporting Information). Remarkably, the results remain consistent regardless of the inclusion of SOC, further affirming that the charge dynamic processes primarily arise from charge transfer rather than spin-flip phenomena.

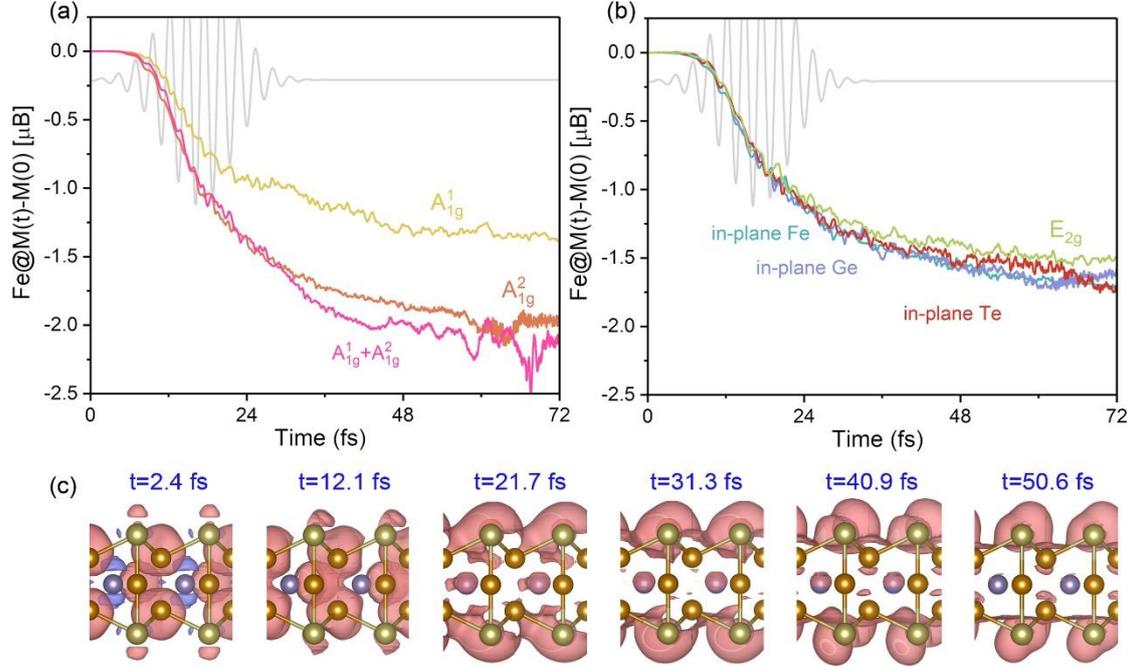

**Figure 4.** (a) Change in the spin moment ($\Delta M(t)=M(t)-M(0)$) dynamics of Fe atom in FGT for full nuclear dynamics with pre-excited $A_{1g}^1$ and $A_{1g}^2$ coherent phonons and when two out-of-plane modes are excited together. (b) The same results of in-plane coherent phonons including $E_{2g}$, in-plane Fe, in-plane Ge and in-plane Te modes. (c) Snapshots of the magnetization density of FGT with pre-excited $A_{1g}^2$ coherent phonon for full nuclear dynamics at different time points. Red (purple) domains of the iso-surface represent spin-up (spin-down) electrons. The iso-surface is set as 0.002 e/Å$^3$.

So far, we have demonstrated the beneficial impact of pre-excited $A_{1g}^1$, $A_{1g}^2$ and $E_{2g}$ coherent phonons on controlling magnetization dynamics, while revealing the underlying physical mechanism behind the increased spin moment loss. To assess the magnetization dynamics of FGT when multiple coherent phonons oscillating parallel to the same spin quantization axis are simultaneously excited, the spin moment loss of Fe atom is explored when out-of-plane $A_{1g}^1$ and $A_{1g}^2$ modes are excited together, as shown in **Figure 4a**. We observed that the spin moment loss of Fe atoms for co-excited $A_{1g}^1$ and $A_{1g}^2$ is close to that for $A_{1g}^2$ mode alone, suggesting that the mode exerting the strongest influence on magnetization will predominantly govern the magnetization dynamics. In addition, the in-plane $E_{2g}$ coherent phonon can be regarded as a collection of in-plane Fe mode, in-plane Ge mode and in-plane Te modes. Consequently, we examine the spin moment loss for these three modes respectively, as shown in Figure 4b. In contrast to the out-of-plane $A_{1g}^1$ and $A_{1g}^2$ modes, the discrepancy between these three in-plane modes is negligible. Therefore, we propose experimentally manipulating magnetization dynamics by selectively pre-exciting out-of-plane coherent phonons, as they exhibit higher sensitivity to controlling spin moment loss compared to in-plane phonons. To gain further insights into spin transfer processes, we analyze the time evolution of magnetization density in FGT with pre-excited $A_{1g}^1$, $A_{1g}^2$ and $E_{2g}$ modes, as shown in Figure 4c and Figure S7 (Supporting Information). Notably, a greater transfer of spin moment from Fe atoms to adjacent Ge/Te atoms is observed in FGT with pre-excited $A_{1g}^2$ mode after ~21 fs. Here the multiple coherent phonon excitations with phase difference as well as non-zero amplitude vector angle, which could form chiral phonons, are not considered in this work. Some specific pulse light sources, e.g. circularly polarized light, can be employed to excite chiral phonons, potentially inducing more exotic photo-induced magnetism and optical phenomena,[41–44] and are worthy of further exploration in the future.

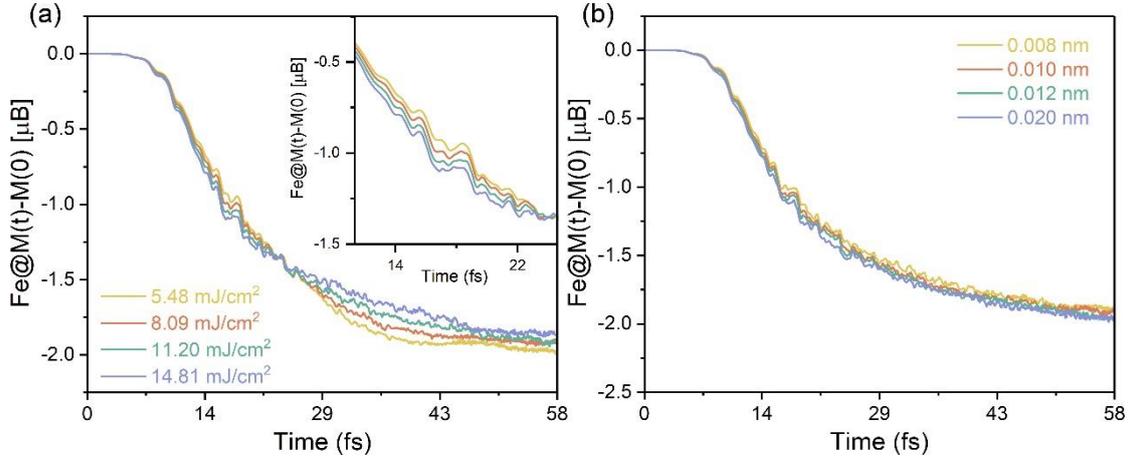

**Figure 5.** Change in the spin moment ($\Delta M(t)=M(t)-M(0)$) dynamics of Fe atom in FGT for full nuclear dynamics with (a) different fluence (FWHM=12.21 fs) and (b) different amplitudes of pre-excited $A_{1g}^2$ phonon.

Next, we examine the $A_{1g}^2$ coherent phonon as a case and investigate the spin moment dynamics of Fe atoms in FGT under the influence of laser pulses with different pulse fluence, as shown in **Figure 5a**. The results reveal that while the spin moment loss exhibits low sensitivity to pulse fluence, a relative dependence is still evident. First, the spin moment loss experiences a slight increase with rising pulse fluence before 24 fs (inset panel in Figure 5a), attributed to the OISTR phenomenon. Following the dissipation of the laser pulse after 24 fs, the spin moment loss shows the opposite trend compared to the earlier time scale. This behavior can be understood as the Fe atom nuclei exhibiting larger adiabatic displacements towards equilibrium with increasing pulse fluences. As a result, weaker interactions between nuclei and electronic systems occur, leading to a reduction in spin moment loss over longer time scales. This suggests that if the fluence threshold is sufficiently strong to restore the Fe atom nuclei from their pre-excited positions to equilibrium, the loss of spin moment will resemble scenarios where nuclear dynamics are not considered. Besides, we investigate the impact of phonon amplitude on magnetization dynamics as well. Figure 5b illustrates that there is a negligible difference in the amount of spin moment loss when the amplitude of phonon varies from 0.008 to 0.020 nm. Hence, we conclude that in the double-pump technique, the initial optical laser pulse solely excites coherent phonons without consideration for phonon amplitudes, whereas the subsequent optical laser pulse plays a certain role in manipulating magnetization dynamics.

## 3. Conclusion

In summary, we have investigated and proposed a coherent phonon-induced control strategy for the magnetization dynamics of 2D ferromagnet Fe$_3$GeTe$_2$ by performing the rt-TDDFT coupled with Ehrenfest dynamics simulations. Our results indicate that selectively pre-exciting coherent phonons will lead to an increase in the spin moment loss of Fe, Ge, and Te atoms in FGT, with the maximum increment reaching up to 53% when the out-of-plane $A_{1g}^2$ coherent phonon is pre-excited. Further spin-resolved charge dynamics analysis reveals that the change in charge of Fe, Ge and Te atoms is predominantly influenced by the nuclei motion of pre-excited coherent phonons over a longer time scale. This will lead to an asymmetric interatomic charge transfer pathway after the laser pulse disappears, thereby manipulating the additional spin moment loss in FGT. Moreover, we explore the spin moment loss of Fe atoms when multiple coherent phonons oscillating parallel to the same spin quantization axis are simultaneously pre-excited. The results reveal that the magnetization dynamics of FGT are sensitive to the out-of-plane instead of in-plane coherent phonon. Additionally, we observe that the magnetization dynamics exhibit a proportionate relationship with pulse fluence in the early stages, while becoming inversely proportional to pulse fluence upon the disappearance of the laser pulse. This is attributed to the larger adiabatic atom displacements towards the equilibrium with the increase of pulse fluences, resulting in weaker interactions between nuclei and electronic systems. In light of these novel results, we suggest using a laser pulse with moderate pulse fluence to selectively pre-excite out-of-plane coherent phonons in experiments to effectively manipulate the magnetization dynamics in 2D magnets.

## 4. Computational Theory and Details

For investigating the dynamics of spin and charge, we have employed state-of-the-art ab initio TDDFT methods.[45] TDDFT systematically transforms the computationally challenging problem involving electron interactions into solving the Kohn-Sham (KS) equation for noninteracting Fermions within an artificial potential. The time-dependent KS equation is:

$$i\frac{\partial \psi_j(\mathbf{r},t)}{\partial t} = \left[\frac{1}{2}\left(-i\nabla + \frac{1}{c}\mathbf{A}_{ext}(t)\right)^2 + v_s(\mathbf{r},t) + \frac{1}{2c}\boldsymbol{\sigma}\cdot\mathbf{B}_s(\mathbf{r},t) + \frac{1}{4c^2}\boldsymbol{\sigma}\cdot[\nabla v_s(\mathbf{r},t)\times -i\nabla]\right]\psi_j(\mathbf{r},t) \quad (1)$$

where $\mathbf{A}_{ext}(t)$ and $\boldsymbol{\sigma}$ represent the vector potential and Pauli matrices. The KS effective potential $v_s(\mathbf{r},t) = v_{ext}(\mathbf{r},t) + v_H(\mathbf{r},t) + v_{xc}(\mathbf{r},t)$ can be decomposed into the external potential $v_{ext}$, the classical Hartree potential $v_H$, and the exchange-correlation (XC) potential $v_{xc}$, respectively. The KS magnetic field can be written as $\mathbf{B}_s(\mathbf{r},t) = \mathbf{B}_{ext}(\mathbf{r},t) + \mathbf{B}_{xc}(\mathbf{r},t)$, where $\mathbf{B}_{ext}$ and $\mathbf{B}_{xc}$ may represent the magnetic field of the applied laser pulse plus an additional magnetic field and XC magnetic field, respectively. The last term in Eq. (1) stands for spin-orbit coupling (SOC).

The back-reaction of the nuclear dynamics on the electronic system is approximated with

$$v_s(\mathbf{r},t) \to v_s(\mathbf{r},t) - \sum_{pa}\frac{\partial v_{cl}(\mathbf{r})}{\partial u_\alpha^p}\delta u_\alpha^p(t) \quad (2)$$

where $p$ labels a nucleus and α is the direction. $v_{cl}$ is the Coulomb potential from the nucleus and the core density of electrons. $\delta u_\alpha^p(t)$ is the displacement of the nucleus away from equilibrium and is calculated from the atomic force.[46]

The structural optimization was implemented with the Vienna Ab initio Simulation Package (VASP).[47,48] The Perdew-Burke-Ernzerhof (PBE) functional within the generalized gradient approximation was employed to account for exchange-correlation interactions.[49] Electron-ion interaction was described using the projector-augmented wave method.[50] A cutoff energy of 500 eV and a Monkhorst-Pack 15×15×1 k-mesh grid were utilized. The lattice constants and atomic positions were fully relaxed until the atomic forces were smaller than 0.1 meV Å$^{-1}$. The electron relaxation convergence criterion was $10^{-7}$ eV. A vacuum region of more than 15 Å in the z direction was used to avoid spurious interactions with the neighboring cells. Phonon calculations were conducted using the finite difference scheme implemented in Phonopy software,[51] adopting a finite-displacement approach with a 0.01 Å displacement and a 4×4×1 supercell.

Laser-induced spin dynamics calculations were implemented with the ELK code[52] using a fully noncollinear version of rt-TDDFT. A regular mesh in a k-space of 8×8×1, a smearing width of 0.027 eV, and a time step of $\Delta t=0.1$ a.u. were used. All calculations were performed using an adiabatic local spin density approximation (ALSDA). The external optical field was defined as a linearly polarized pulse with a photon energy of 1.55 eV, full width at half maximum (FWHM) of 12.21 fs, and a pulse fluence of 11.20 mJ/cm$^2$. TDDFT coupled with Ehrenfest nuclear dynamics simulation involves a two-step process: First, the forces under the influence of a laser pulse are calculated. In the second step, the same calculation is repeated but this time the forces are used to find the motion of the nuclei. The back-reaction of this nuclear motion is applied to the electronic system.

**Acknowledgements**

We acknowledge the e-INFRA CZ (ID:90140) for providing computational resources. J. He acknowledged the support from MSCA Fellowships CZ−UK with CZ.02.01.01/00/22_010/0002902. Z. Zhou acknowledged the support from the JUNIOR Fund of Charles University in Prague.

**Conflict of Interest**

The authors declare no conflict of interest.

**Data Availability Statement**

The data that support the findings of this study are available from the corresponding author upon reasonable request.

Supporting Information for

# Coherent Phonon Control of Ultrafast Magnetization Dynamics in Fe$_3$GeTe$_2$ from Time-Dependent *Ab Initio* Theory

Zhaobo Zhou[1], Min Li[1], Thomas Frauenheim[2,3], Junjie He[1,*]

[1]Department of Physical and Macromolecular Chemistry, Faculty of Science, Charles University in Prague, Prague 12843, Czech Republic
[2]School of Science, Constructor University, Bremen 28759, Germany
[3]Institute for Advanced Study, Chengdu University, Chengdu 610106, China
*E-mail: junjie.he@natur.cuni.cz

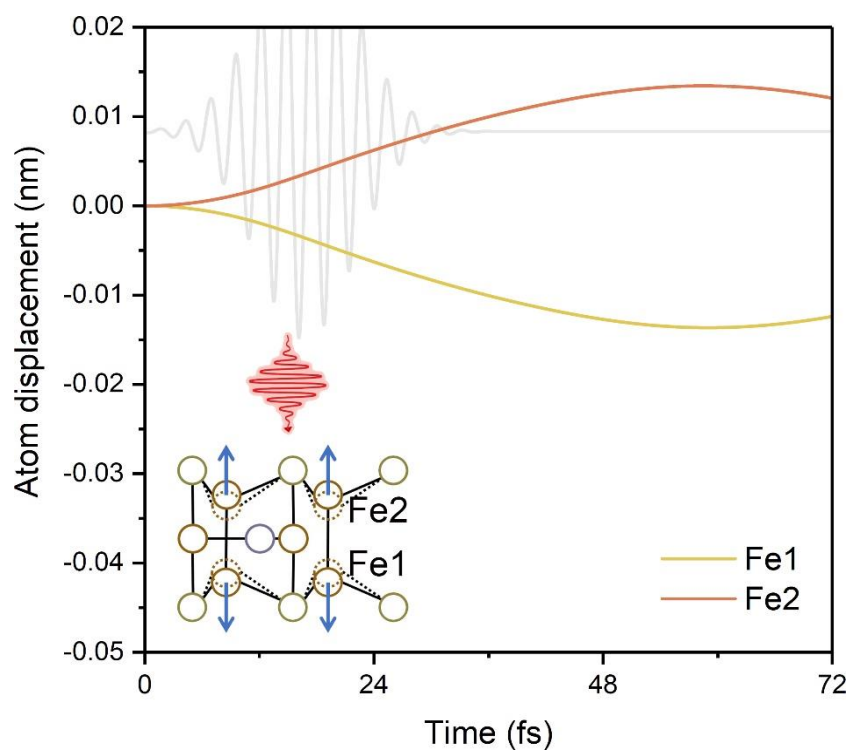

**Figure S1.** The atom displacement of Fe1 and Fe2 atoms in FGT during the pre-excited phonon mode. The linearly polarized laser pulse (grap line) with a photon energy of 1.55 eV, a 12.21 fs FWHM and 11.20 mJ/cm$^2$ along the out-of-plane direction is applied. The maximum displacement is 0.013 nm with a time period of 60 fs.

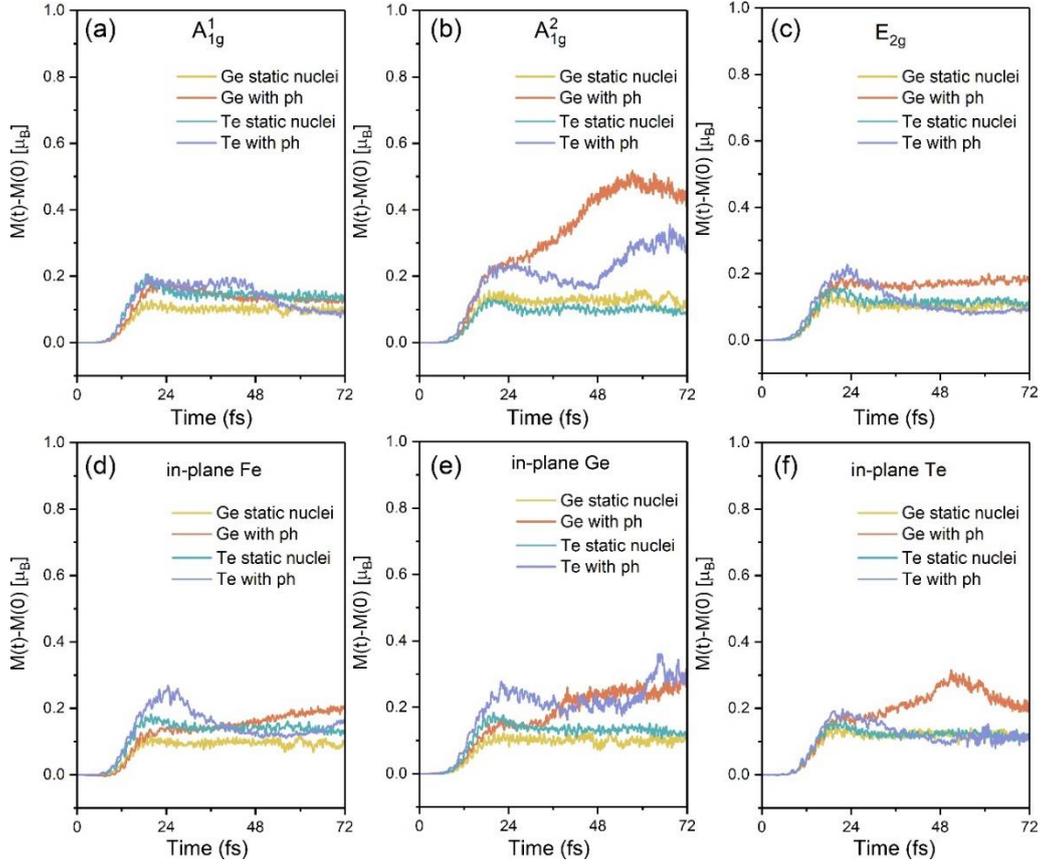

**Figure S2.** Change in the spin moment ($\Delta M(t)=M(t)-M(0)$) dynamics of Ge and Te atoms in FGT for full nuclear dynamics (Ge/Te with ph) and in the absence of nuclear dynamics (Ge/Te static nuclei). Results are shown for six pre-excited coherent phonon modes: (a) $A_{1g}^1$, (b) $A_{1g}^2$, (c) $E_{2g}$, (d) in-plane Fe, (e) in-plane Ge and (f) in-plane Te.

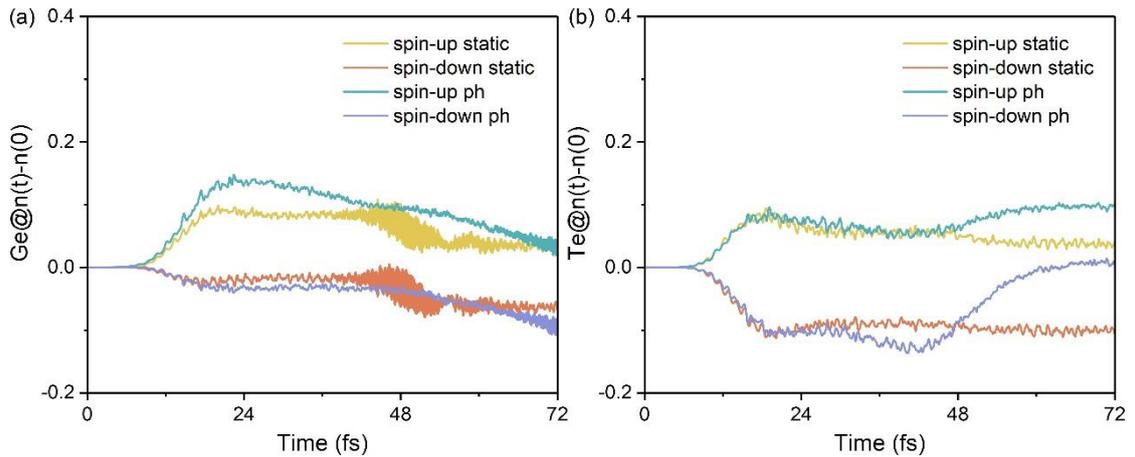

**Figure S3.** Change in the spin-resolved charge ($\Delta n(t)=n(t)-n(0)$) of (a) Ge and (b) Te atom in FGT with (static) and without (ph) nuclear dynamics. Results are shown for pre-excited $A_{1g}^1$ coherent phonons. The positive (negative) value means the increase (decrease) of electrons.

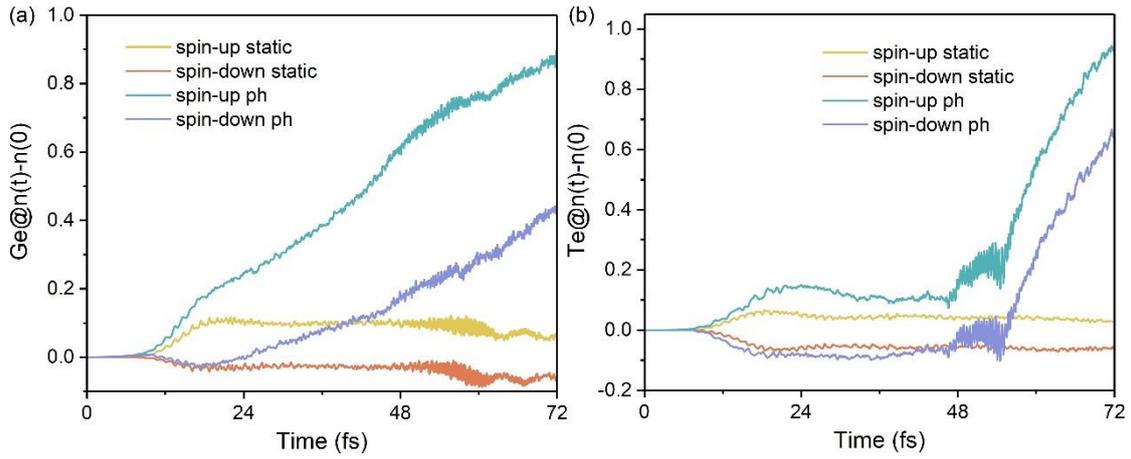

**Figure S4.** Change in the spin-resolved charge (Δn(t)=n(t)-n(0)) of (a) Ge and (b) Te atom in FGT with (static) and without (ph) nuclear dynamics. Results are shown for pre-excited $A_{1g}^2$ coherent phonons. The positive (negative) value means the increase (decrease) of electrons.

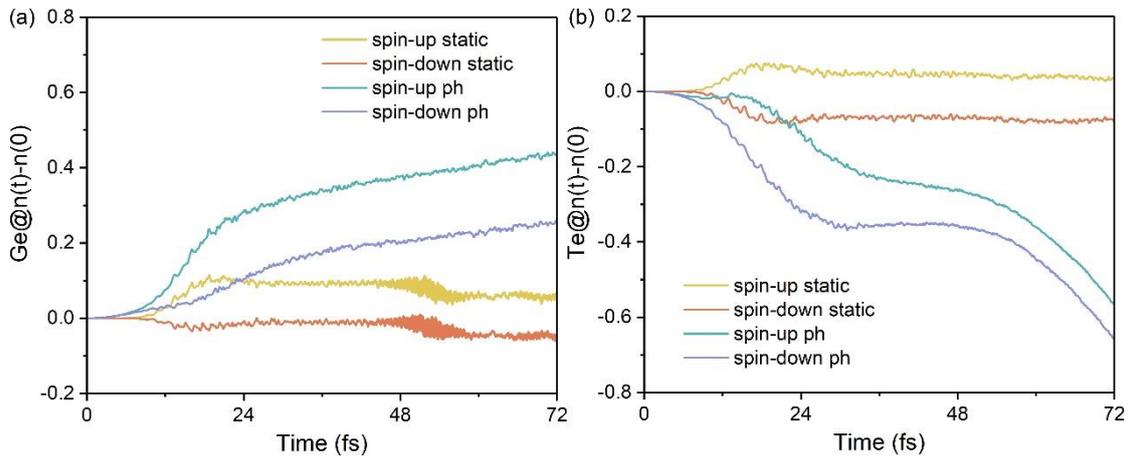

**Figure S5.** Change in the spin-resolved charge (Δn(t)=n(t)-n(0)) of (a) Ge and (b) Te atom in FGT with (static) and without (ph) nuclear dynamics. Results are shown for pre-excited $E_{2g}$ coherent phonons. The positive (negative) value means the increase (decrease) of electrons.

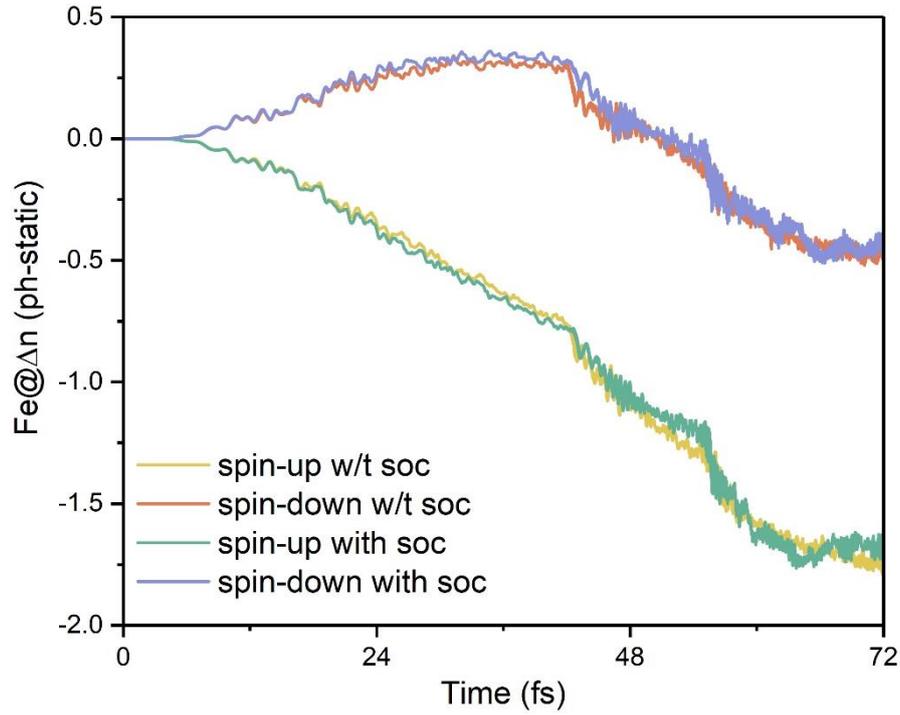

**Figure S6.** Charge difference of Fe atom between full nuclear dynamics and in the absence of nuclear dynamics of pre-excited $A_{1g}^2$ coherent phonons. The results with and without SOC are calculated. The positive (negative) value means the increase (decrease) of charge.

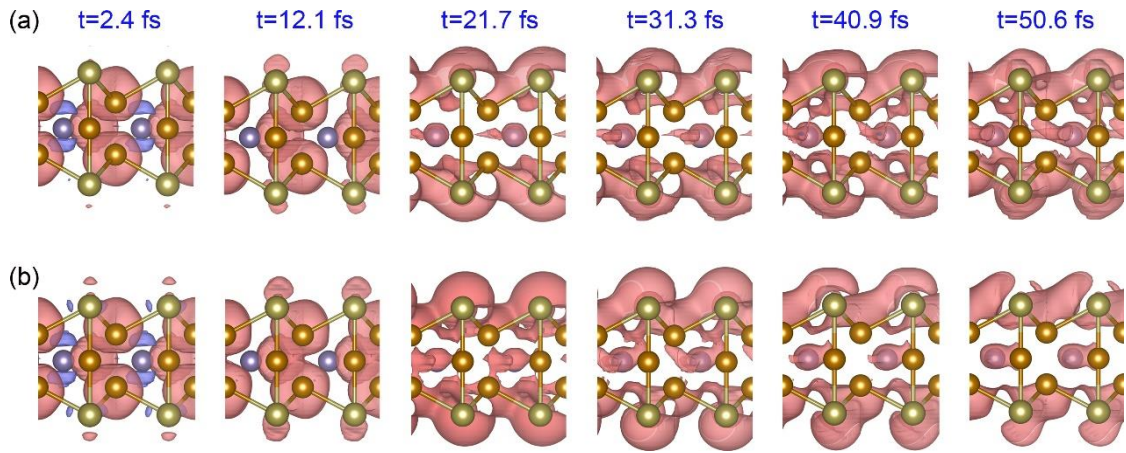

**Figure S7.** Snapshots of the magnetization density of FGT for full nuclear dynamics with pre-excited (a) $A_{1g}^1$ and (b) $E_{2g}$ coherent phonon at different time points. Red (purple) domains of the iso-surface represent spin-up (spin-down) electrons. The iso-surface is set as 0.002 $e/Å^3$.